\documentstyle[a4,12pt]{article} 

\renewcommand{\gg}{\gamma} 
\newcommand{\gd}{\delta} 
\renewcommand{\ge}{\epsilon}

\newcommand{\gvf}{\varphi}

\newcommand{\gm}{\mu}

\newcommand{\gk}{\kappa} 
\newcommand{\gl}{\lambda} 
 
\newcommand{\gth}{\theta} 
\newcommand{\gs}{\sigma}

\newcommand{\gps}{\psi}

\newcommand{\gG}{\Gamma} 
\newcommand{\gD}{\Delta} 
\newcommand{\gF}{\Phi}

\newcommand{\gL}{\Lambda}

\newcommand{\gO}{\Omega} 
 
\newcommand{\gPs}{\Psi} 
 
\newcommand{\cL}{{\cal L}} 
\newcommand{\cM}{{\cal M}}

\newcommand{\uI}{{\underline I}} 
\newcommand{\uJ}{{\underline J}} 
\newcommand{\uK}{{\underline K}}

\newcommand{\ba}{{\bar a}}

\newcommand{\bz}{{\bar z}} 
\newcommand{\bA}{{\bar A}} 
\newcommand{\bB}{{\bar B}}

\newcommand{\bF}{{\bar F}} 
 
\newcommand{\bH}{{\bar H}}

\newcommand{\bZ}{{\bar Z}} 
\newcommand{\bge}{{\bar\epsilon}}

\newcommand{\bgvf}{{\bar\varphi}}

\newcommand{\bgl}{{\bar\lambda}} 
 
\newcommand{\bgth}{{\bar\theta}}

\newcommand{\bgps}{{\bar\psi}}

\newcommand{\bgF}{{\bar\Phi}}

\newcommand{\bgPs}{{\bar\Psi}} 
 
\newcommand{\slashed}{\hspace{-1.1ex}/} 
\newcommand{\Slashed}{\hspace{-1.5ex}/\hspace{.6ex}}

\newcommand{\der}{\partial} 
\newcommand{\dd}[2]{\frac{\der #1}{\der #2}}
\newcommand{\Der}{D} 
\newcommand{\sDer}{\Der\Slashed} 
\newcommand{\sder}{\der\slashed}

\newcommand{\nit}{\noindent} 
\newcommand{\nl}{\newline} 
\newcommand{\np}{\newpage} 
\newcommand{\dsp}{\displaystyle}

\newcommand{\ct}{\cite} 
\newcommand{\bit}{\bibitem} 
\newcommand{\lh}{\left(} 
\newcommand{\rh}{\right)} 
\newcommand{\ld}{\left.} 
\newcommand{\rd}{\right.}

\newcommand{\vs}[1]{\vspace{#1 ex}}  
\newcommand{\hs}[1]{\hspace{#1 em}}
\newcommand{\labl}[1]{\label{#1}} 
\newcommand{\Kh}{K\"{a}hler}  
\newcommand{\beq}{\begin{equation}} 
\newcommand{\feq}{\end{equation}} 
\newcommand{\barr}{\begin{array}} 
\newcommand{\earr}{\end{array}}

\newcommand{\non}{\nonumber} 

\begin{document} 

\pagestyle{empty} 

\begin{flushright}
NIKHEF 2003-002 \\
UVIC-TH/02-03~~~
\end{flushright} 
\vs{10} 

\begin{center} 
{\large {\bf Singular supersymmetric $\gs$-models}} \\
\vs{7} 

T.S.\ Nyawelo$^a$, F.\ Riccioni$^{b}$, J.W.\ van Holten$^c$ 
\vs{2} 

NIKHEF\\
PO Box 41882 \\
1009 DB Amsterdam  \\ 
The Netherlands \\
\vs{3}

S.\ Groot Nibbelink$^d$ \\
(CITA National Fellow)
\vs{2} 

Univ.\ of Victoria, Dept.\ of Physics and Astronomy \\ 
PO Box 3055 STN CSC \\
Victoria BC, V8W 3P6 \\
Canada\\
\vs{3}
Febr.\ 2, 2003

\end{center} 

\nit
{\small 
{\bf Abstract} \\
Supersymmetric non--linear $\gs$--models are described by a field 
dependent \Kh\ metric determining the kinetic terms. In general 
it is not guaranteed that this metric is always invertible. 
Our aim is to investigate the symmetry structure of supersymmetric 
models in four dimensional space-time in which metric singularities 
occur. For this purpose we study a simple anomaly-free extension of 
the supersymmetric $CP^1$ model from a classical point of view.
We show that the metric singularities can be regularized by 
the addition of a soft supersymmetry-breaking mass parameter. 
}

\vfill 

\nit
\footnoterule 
\nit
{\footnotesize{$^a$ e-mail: tinosn@nikhef.nl \\
$^{b}$ e-mail: fabio@nikhef.nl \\
$^c$ e-mail: v.holten@nikhef.nl \\ 
$^d$ e-mail: grootnib@uvic.ca }}

\np 
~\hfill 

\np

\pagestyle{plain} 
\pagenumbering{arabic} 

\nit

\nit
{\bf 1.\ Introduction} 
\vs{1}

\nit
The scalar fields of supersymmetric $\gs$--models in four dimensions
take values in K\"{a}hler manifolds \ct{zumino,freed-alg}. In some
supersymmetric field theories the \Kh\ metric develops a zero mode; 
then the model becomes singular in the sense that some of the kinetic 
terms vanish in the vacuum state, and correspondingly some couplings 
diverge. The central issue we want to investigate in this paper is how to
treat supersymmetric field theories in which these types of complications  
occur. This analysis is in particular relevant for supersymmetric non-linear 
$\gs$--model building based on homogeneous \Kh ian cosets $G/H$ 
\ct{sgn-n-vh}. 

Supersymmetric pure $\gs$--models on cosets, including among 
others the grassmannian models on $SU(n+m)/[SU(n) \times SU(m) 
\times U(1)]$ and the models on manifolds $SO(2n)/U(n)$, are 
known to be anomalous \ct{moor-nel,coh-gom,buch-ler}, since they 
incorporate chiral fermions in non-trivial representations of the 
holonomy group. These anomalies can be removed  
by coupling additional chiral superfields carrying specific
line-bundle representations of the group $G$ \ct{sgn-vh,sgn}. 
In a study of anomaly-free extended supersymmetric $\gs$--models 
on coset manifolds of the type $SO(2n)/U(n)$, it was 
found that the resulting field metric can develop singularities in the
form of zero-modes, and therefore may not be positive definite
 \ct{sgn-n-vh}. Indeed, upon gauging all or part of the isometry group the 
$D$-term potential sometimes forces the scalar fields to take 
vacuum expectation values at the geometric singularities of field
space. 

In order to gain an understanding of this situation, in this paper we 
study an anomaly-free extension of the $d = 4$ supersymmetric 
$CP^1$-model, where the scalar fields take values in $SU(2)/U(1)$, 
and some of the isometries are gauged. In addition to the chiral multiplet 
parametrizing the coset manifold, anomaly cancellation requires the 
inclusion of other chiral multiplets. The simplest choice corresponds 
to a single supermultiplet with the scalar component defining a section 
of a complex line bundle. We couple these matter multiplets to a gauge 
multiplet, focussing in particular on the case where the full $SU(2)$ 
isometry group is gauged. If one considers the most general \Kh\ potential
corresponding to this geometry, one realizes that depending on the 
parameters, the resulting metric can vanish for particular values of 
the scalars. Moreover, in many cases the potential drives the scalars 
to a vacuum value exactly at these singular points. At this singularity, 
some of the four-fermi couplings explode, while the mass terms for 
the fermions stay in general finite. 

Singularities can occur in two places: either the kinetic term of
the scalar parametrizing $CP^1$ or the kinetic term of the scalars 
parametrizing the section of the line bundle can vanish in the vacuum. 
When the two singularities occur at the same point, the vacuum 
preserves both supersymmetry and the whole $SU(2)$ gauge symmetry. 

The paper is organized as follows. In section 2 we describe the isometry 
structure of the scalar manifold, and we show how a generic choice of the 
\Kh\ potential leads to geometrical singularities. In Section 3 we
describe the gauged version of the anomaly-free $CP^1$-model. In section 
4 we classify the possible vacua, discussing general consequences for the 
gauge symmetries and particle spectra. Section 5 discusses a modification 
of the model containing a soft supersymmetry breaking mass term which 
preserves the full non-linear $SU(2)$. The mass term acts as a regulator, 
as it displaces the vacuum away from the singular point. The particle 
spectrum in this regulated model is computed and shown to be 
sensitive to the behaviour of the K\"{a}hler potential in the limit of 
small regulator mass.  
In section 6 we present some examples, showing that the various
types of behaviour of the spectra in the limit of small regulator 
mass can all be realized in actual models.  In section 7 we 
summarize our results. In the two appendices we present our notation 
and we give the complete expressions for the lagrangeans both 
off-shell and in the unitary gauge. 
\nl

\nit
{\bf 2.\ The model} 
\vs{1}

\nit
We consider a simple supersymmetric $\gs$-model, in which the 
vanishing of kinetic terms occurs in a restricted subset of the 
parameter domain. The model is based on the supersymmetric $CP^1$-model, 
where the scalar fields take values in $SU(2)/U(1)$. As the pure 
supersymmetric $CP^1$ model in four dimensions is anomalous, we 
include another chiral multiplet, transforming as a contravariant 
vector on the $CP^1$ manifold. The complete field content of the 
model is therefore specified by a complex scalar superfield $\gF 
= (z, \psi_L, H)$ and a second complex scalar superfield $A = (a, 
\gvf_L, B)$. These superfields define representations of the 
isometry group $SU(2)$; on the scalar fieldsthey take the 
infinitesimal form 
\beq 
\gd z = \ge + i \gth z + \bar{\ge} z^2, \hs{2} 
\gd a = - i \gth a - 2 \bar{\ge} z a.
\label{1}
\feq 
Here $\gth$ is the parameter of $U(1)$ phase transformations, and 
$(\ge, \bar{\ge})$ are the complex parameters of the broken off-diagonal 
$SU(2)$ transformations. We take the fields $z$ and $a$ to be dimensionless;
dimensionful fields are obtained by introducing a parameter $f$ with the 
dimension of inverse mass (in natural units in which $c = \hbar = 1$), and 
making the replacements 
\beq 
z\, \rightarrow\, fz, \hs{2} a\, \rightarrow\, fa,
\label{1.0}
\feq 
and similarly for other fields to be introduced. The infinitesimal transformations 
of such other field components (chiral spinors, auxiliary fields) are found by 
requiring the isometries to commute with supersymmetry \ct{jwvh}. Observe, that 
the opposite linear $U(1)$ transformations of the multiplets are precisely as 
required for cancellation of the isometry anomalies. 

The dimensionless $CP^1$ K\"{a}hler potential 
\beq
K_{\gs}(\bz,z) = \ln (1 + \bz z)
\label{2}
\feq
is invariant under the isometry transformations (\ref{1}) up
to the real part of a holomorphic function:
\beq 
\gd K_{\gs}(\bz,z) = F(z) + \bF(\bz), \hs{2} 
 F(z;\gth,\bge) = \frac{i}{2}\, \gth + \bge z.
\label{3}
\feq 
The transformations of the scalar $a$ can therefore be rewritten as 
\beq 
\gd a = - 2 F(z;\gth,\bge) a.
\label{4}
\feq
It follows, that the dimensionless real scalar 
\beq 
X = \ba a\, e^{2K_{\gs}(\bz,z)}= \ba a\, (1 + \bz z)^2,
\label{5}
\feq 
is an invariant under the full set of isometries. With this observation 
in mind, we take as the starting point for our supersymmetric model a 
K\"{a}hler potential
\beq
K(\bgF,\gF;\bA,A) = K_{\gs}(\bgF,\gF) + K_m(\bgF,\gF;\bA,A) = 
 \ln (1 + \bgF \gF) + K_m(\gO),
\label{6}
\feq 
with $K_m(\gO)$ a real function of the real superfield
\beq 
\gO = \bA A\, e^{2 K_{\gs}(\bgF,\gF)}, 
\label{7}
\feq 
of which the real scalar quantity $X$ is the lowest component. 
The kinetic terms of scalars and chiral spinors in the action 
are then given by 
\beq
S = \int d^4x \int d^2 \gth d^2\bgth\, K(\bgF,\gF;\bA,A)
 = - \int d^4x\, G_{I\uI} \lh \der \bZ^{\uI} \cdot \der Z^I + 
 \bgps_L^{\uI} \stackrel{\leftrightarrow}{\sDer} \gps_L^I \rh + ...
\label{8}
\feq
Here $Z^I = (z,a)$ and $\gps^I_L = (\gps_L, \gvf_L)$, $I = (z,a)$, 
denote the scalar and spinor components of the respective superfields 
$\gF^I = (\gF, A)$, and the dots represent four-fermion interactions. 
The full component action and its derivation are presented in appendices 
A and B. The K\"{a}hler metric in field space is obtained from the K\"{a}hler 
potential: 
\beq 
\barr{l} 
G_{I\uI}= \dsp{ \frac{\der^2 K}{\der Z^I \der\bZ^{\uI}} 
 = \lh \barr{cc}  \dsp{
\frac{2 M(X) + 4\bz z X M^{\prime}(X)}{(1+ \bz z)^2}  }
  &  2 a \bz (1 +\bz z)\, M^{\prime}(X) \\
 & \\
 2 \ba z (1 + \bz z)\, M^{\prime}(X) & 
 (1 + \bz z)^2 M^{\prime}(X)  \earr \rh, }
\earr 
\label{9}
\feq 
where we have introduced the $SU(2)$-invariant function $M(X)$ 
defined in terms of $K_m(X)$ as 
\beq 
M(X) = \frac{1}{2}\, + X K_m^{\prime}(X).
\label{9.1}
\feq
The primes in the equations denote derivatives w.r.t.\ $X$. 
The  determinant of the metric is 
\beq 
\det G_{I\uI} = 2 M^{\prime}(X) M(X). 
\label{10}
\feq 
Positive definite kinetic terms are obtained if both
\beq
 M(X)  > 0, \hs{2} M^{\prime}(X) > 0.
\label{9.2}
\feq
We then have a standard non-linear field theory, which is well-behaved
below a cut-off $\gL^2 \sim {\cal O}(1/f^2)$, the parameter 
determining the characteristic scale of the K\"{a}hler manifold. 

In contrast, if one of the two factors is negative: $\det G < 0$, the theory 
contains ghosts and is inconsistent. If one of the two factors vanishes, 
we have critical case and we must resort to some regularization in order
to investigate if the model still allows for some physically interesting  
interpretation. Observe however, that with the given field content it is not 
possible to construct an $SU(2)$-invariant superpotential $W(\gF^I)$. 
With such a flat potential the vacuum expectation values of the scalar 
fields are undetermined, but it is natural to suppose they are to be fixed 
in the region where the model is well behaved according to the criterion 
of eq.\ (\ref{9.2}). 

It is important to stress that the vanishing of the kinetic terms 
for the scalar fields corresponds to the divergence of some four-fermi 
couplings in the lagrangian, once one solves the equations for the auxiliary
fields (see Appendix A for the details). 
The four--fermion interactions take the general form  
\beq
\barr{lll}
L_{4fermi} = 
K_{z \bz z \bz}\,  \bgps_R \gps_L \,  \bgps_L \gps_R 
+ 
K_{a \ba a \ba}\,  \bgvf_R \gvf_L \,  \bgvf_L \gvf_R 
\non \\
\quad \\
\qquad \quad \ + \big\{  
K_{z \bz a \ba} \,  \bgps_R \gps_L \,  \bgvf_L \gvf_R 
+ {\rm perm.}
\big\} 
\ , 
\earr
\labl{4fermi} 
\feq
with the curvature components given by  
\beq
\barr{lll}
K_{z \bz z \bz} =  -4 M + 8 M' X  
\ , \\ 
\quad \\
K_{a \ba a \ba} = 
M'' + X M'''  - X \frac { {M''}^2}{ M'}  \ , 
\\
\quad \\
K_{z \bz a \ba} =  
2 M' + 2 X M''  - 2X \frac { {M'}^2}{ M} \ . 
\earr
\label{4fermicouplings}
\feq
The last term in eq. (\ref{4fermi}) is short-hand for four combinations 
with two $\gps$'s and two $\gvf$'s which contribute to the four-fermion 
terms. From the expressions (\ref{4fermicouplings}) it follows, that 
subsets of the four-fermion terms diverge at the kinetic singularities.
\nl

\nit
{\bf 3.\ The gauged $CP^1$-model} 
\vs{1} 

\nit
The critical case is of importance when the isometry group 
of the model is gauged. As well-known, the gauging of the supersymmetric 
model involves several steps: \nl
- modification of the kinetic terms by introducing gauge-covariant
derivatives; \nl
- addition of a $D$-term potential; \nl
- addition of Yukawa couplings for the fermions; \nl
- introduction of kinetic terms for the gauge superfields. \nl
A simplification occurs however, because in the model with broken local 
$SU(2)$  the Goldstone bosons are absorbed completely in the longitudinal
components of the massive charged vector bosons, and we can analyze the
model in the unitary gauge $\bz = z =0$, with $X = \ba a$. 
In this gauge the metric (\ref{9}) is automatically diagonal:
\beq 
G_{I\uI} = \lh \barr{cc} 2 M(X)  & 0 \\
                         0 & M^{\prime}(X) \earr \rh.
\label{14}
\feq
The expressions for the gauge-covariant derivatives of the complex scalar 
fields read
\beq 
\barr{lll}
D_{\mu} z & = & \dsp{ 
\der_{\mu} z -  ig A_{\mu} z - \frac{g}{\sqrt{2}} \lh W^+_{\mu}  
 + W^-_{\mu} z^2  \rh
 \simeq - \frac{g}{\sqrt{2}}\, W^+_{\mu}, }\\
 & & \\
D_{\mu} a & = & \der_{\mu} a + ig A_{\mu} a + \sqrt{2}\, g W^-_{\mu} z a\,  
 \simeq\, \der_{\mu} a + ig A_{\mu} a, 
\earr 
\label{15}
\feq
whilst the covariant derivatives of the fermions become
\beq
\barr{lll}
D_{\mu} \psi_L & = & \der_{\mu} \psi_L - ig A_{\mu} \psi_L - \sqrt{2}\, g W^-_{\mu}
 z \psi_L \simeq \der_{\mu} \psi_L - ig A_{\mu} \psi_L \\
 & & \\
D_{\mu} \gvf_L & = & \der_{\mu} \gvf_L + ig A_{\mu} \gvf_L + 
 \sqrt{2}\, g W^-_{\mu} (z \gvf_L + a \psi_L) \\
 & & \\ 
 & \simeq & \der_{\mu} \gvf_L + ig A_{\mu} \gvf_L + 
 \sqrt{2}\, g W^-_{\mu} a \psi_L.
\earr 
\label{15.a}
\feq
The last expression on each line is the one in the unitary gauge. We have 
introduced the notation $W^{\pm}_{\mu}$ for the charged gauge fields corresponding 
to the broken $SU(2)$ transformations parametrized by $(\ge, \bge)$; $A_{\mu}$ is 
the gauge field of the $U(1)$ transformations.  
Note, that for the unitary gauge to be valid, we must assume  that the charged 
vector bosons are massive, i.e.\ $\langle M(X) \rangle > 0$. In the critical 
case $\langle M(X) \rangle = 0$ the charged vector bosons become massless 
again, and the choice of  the unitary gauge is not allowed. 

Next we discuss the $D$-term potential. 
First we recall, that the isometries (\ref{1}) can be obtained locally 
as gradients of a set of real Killing potentials, defined by 
\beq 
\cM(\gth,\ge,\bge) =  \frac{\gth ( 1 - \bz z) + 
 2 i (\ge \bz - \bge z)}{1 + \bz z}\, M(X). 
\label{11}
\feq
Indeed, the variations (\ref{1}) are given by
\beq
\gd Z^I =  -i G^{I\uI}\, \dd{M}{\bZ^{\uI}}.
\label{12}
\feq 
Now the auxiliary $D$-fields couple to these Killing potentials, and 
after elimination of the $D$-fields the potential for the model with 
fully gauged $SU(2)$ becomes
\beq
V_D = \frac{g^2}{2}\, \dd{\cM}{\ge} \dd{\cM}{\bge}\, + \frac{g^2}{2}\, 
 \lh \dd{\cM}{\gth} \rh^2 = 
 \frac{g^2}{2}\,  M^2(X). 
\label{13}
\feq 
Finally, we also have to introduce kinetic terms for the gauge fields.
They are of the canonical form
\beq
\cL_{gauge} = - \frac{1}{2}\, F^+(W) \cdot F^-(W) - \frac{1}{4}\, F^2(A).
\label{3.1}
\feq
In this expression the quadratic terms for the auxiliary $D$-fields have been 
left out, their elimination giving rise to the scalar potential (\ref{13}). 
For the full action, including the fermionic terms required by supersymmetry,
we again refer to appendix B, eqs.\ (\ref{b.2}) and (\ref{b.6}). \nl 

\nit
{\bf 4.\ Analysis of the particle spectrum} 
\vs{1} 

\nit
We begin our analysis of the particle content of the model by studying the 
bosonic part of the $CP^1$-model, which up to the kinetic terms for the 
gauge bosons is described by the action
\beq
\cL_B = - g^2 M(X) W^+ \cdot W^-  - M^{\prime}(X) |Da|^2 - 
 \frac{g^2}{2}\, M^2(X). 
\label{3.2}
\feq
The vacuum expectation value of the scalar field $a$ is derived by minimizing the
potential, which leads to the condition
\beq
M(X) M^{\prime}(X)\, a = 0. 
\label{3.3}
\feq
It is clear that a priori there may be three ways to solve this equation: \nl
(a) $a = 0$ is always a solution. \nl 
(b) There may be a value $a = a_0$ such that $M^{\prime}_0 = 
M^{\prime}(X_0) = 0$; if the potential reaches its minimum here, then the 
model is critical in the sense discussed above, and we have to be careful 
in the analysis of the physical realization of the theory. \nl
(c) The solution $M_0 = M(X_0) = 0$ is also logically allowed; 
it implies that the charged vector bosons become massless. 
It may happen that at the same time $X_0 M^{\prime}(X_0) = 0$; 
then all $SU(2)$ gauge bosons become massless and the full gauge symmetry 
is restored. In that case the complex scalars $(z, \bz)$ are no longer 
Goldstone bosons, and the unitary gauge can not be used to eliminate them. 
We also observe, that if the solution $M_0 = 0$ exists, it is necessarily 
the absolute minimum of the potential: $V_0 = 0$, and supersymmetry is 
apparently restored as well. Of course, the standard way to describe a 
situation in which the gauge symmetry is restored, is to reformulate the 
physics in terms of linear representations of the gauge symmetry. 

Assuming therefore that supersymmetry is broken and $M_0 > 0$, it
depends on the precise form of $M(X)$ which solution is
the true minimium of the potential; actually, it can happen that
both conditions hold simultaneously. Quite generally, we can derive 
the linearized field equation for fluctuations around the vacuum, by 
making the expansion
\beq
a = a_0 + \sqrt{Z_a}\, \gd a,
\label{3.4}
\feq
where the normalization factor $Z_a$ is still to be determined. 
With $a_0$ a solution of eq.\ (\ref{3.3}), the quadratic part of the 
action for the linearized field fluctuations becomes
\beq 
\barr{lll}
\cL_{lin} & = &  \dsp{ -  g^2 M_0 W^+ \cdot W^- 
 - g^2 X_0 M_0^{\prime} A_{\mu}^2 
 - Z_a\, M^{\prime}_0 \,|\der_{\mu} \gd a |^2 - \frac{g^2}{2}\, M_0^2  }\\ 
 & & \\
 & & \dsp{  
 - g^2 Z_a M_0 M_0^{\prime} |\gd a|^2 - \frac{g^2 Z_a}{2}\, 
 \lh M_0 M^{\prime\prime}_0 + M_0^{\prime\, 2} \rh 
 \lh \ba_0 \gd a + a_0 \gd \ba \rh^2 + ... }
\earr
\label{3.5}
\feq
First we consider the solution (a): $a_0 = 0$ with $M^{\prime}(0) > 0$.
Taking $Z_a = 1/M^{\prime}(0)$ we get a canonically normalized model
for the fluctuating field, satisfying in the linearized limit a Klein-Gordon 
equation 
\beq
\lh - \Box  + m_a^2 \rh \gd a = 0,
\label{3.6}
\feq
with the mass given by 
\beq
m_a^2 = g^2 M(0) = \frac{g^2}{2}.
\label{3.7}
\feq
This conclusion can be extended to the case $M^{\prime}(0) = 0$, 
although $Z_a$ diverges in that case, because mass and kinetic terms are
than rescaled by the same infinite normalization factor, and the mass
remains finite. In fact, from eq.(\ref{4fermi}) it turns out that also the
four-fermion terms stay finite in this case. 

In contrast, in case (b) with $X_0 = |a_0|^2 > 0$ and $M_0^{\prime} = 0$, 
some contributions to the mass terms of the scalars $\gd a$ in eq.\
(\ref{3.5}) generally diverge. This indicates that one of the degrees of 
freedom does not describe a propagating particle: it decouples from the 
spectrum of states. In all cases, when $X_0 M_0^{\prime} = 0$, the mass 
of the  $U(1)$ gauge boson vanishes. 
 
Returning to the possibility (a) with unbroken $U(1)$ gauge symmetry, 
the mass spectrum of the bosons is well-defined and can be read off from 
the action (\ref{3.5}); they are summarized in table 1 below: 

\begin{center}
~~~~~~~~~~~~~~~~\begin{tabular}{|l|c|c|c|} \hline 
mass &  $m_a^2$ & $m_A^2$ & $m_W^2$ \\ \hline
value & $g^2 M(0) = g^2/2$ & 0 & $g^2 M(0) = g^2/2$ \\ \hline
\end{tabular} \nl

{\footnotesize{Table 1: Boson mass spectrum for $X_0 = 0$}} 
\end{center} 

\nit
Next we turn to the spectrum of fermions. With $U(1)$ not broken,
the fermions must fall into charged states. A positively charged 
Dirac fermion is formed by the quasi-Goldstone fermion $\psi_L$
and the gaugino $\gl^+_R$:
\beq
\Psi = \sqrt{2M_0}\, \psi_L +  \gl^+_R,
\label{4.1}
\feq
whereas the remaining fermions $\bgvf_L$, $\gl^-_L$ and $\gl_R$
remain massless. The mass of the Dirac fermion (\ref{4.1}) is 
$m_{\Psi}^2 = 2g^2 M(0) = g^2$. A straightforward calculation
shows, that in the scenario with manifest non-broken $U(1)$ the 
standard supertrace formula for the mass spectrum is satisfied:
\beq
\mbox{Str}\, m^2 = \sum_J (-1)^{2J} (2J+1)\, m_J^2 = 0.
\label{4.2}
\feq

\nit
{\bf 5.\ Softly broken supersymmetry} 
\vs{1}

\nit
As eq.\ (\ref{3.3}) with $X_0 > 0$  implies that $M_0 M^{\prime}_0 = 0$, 
the phase with spontaneosuly broken $U(1)$ symmetry is always critical.
The analysis of the theory is then complicated by the appearance of
infinities at the classical level. However, a completely finite theory 
is obtained by adding an $SU(2)$-invariant soft supersymmetry breaking 
scalar mass term $\gD V(X) = - \mu^2 X$ to the potential. In the following 
we take the point of view that the critical model is the limit of this 
regulated theory when the soft mass term is taken to vanish.

With the addition of the regulator mass term, the full potential becomes
\beq
V(X) = V_D(X) + \gD V(X) = \frac{g^2}{2}\, M^2(X) - \mu^2 X.
\label{5.1}
\feq 
As a result the minimum of the potential is shifted to the position where
\beq 
g^2 M^{\prime} M\, a = \mu^2 a. 
\label{5.2}
\feq
Then either $a_0 = 0$ and $U(1)$ is not broken, as discussed previously; 
or $U(1)$ is broken, $X_0 = |a_0|^2 > 0$ and
\beq
g^2 M^{\prime}_0 M_0 = \mu^2. 
\label{5.3}
\feq
Hence the soft supersymmetry breaking term shifts the vacuum of the model 
away from the critical point. Taking the broken $U(1)$ invariance into 
account, we parametrize the complex scalar field $a$ as
\beq
a = \lh \sqrt{X_0} + \sqrt{\frac{Z_h}{2}}\, h \rh 
 e^{i \sqrt{Z_{\gth}/2}\, \gth},
\label{5.4}
\feq 
where again $Z_h$ and $Z_{\gth}$ are normalization constants to be fixed 
such that we obtain canonically normalized kinetic terms. The bosonic terms 
in the action then become in the unitary gauge
\beq
\barr{lll}
\cL_{bos} & = & \dsp{ - g^2 M_0 W^+ \cdot W^- 
 - \frac{Z_h M_0^{\prime}}{2}\, (\der _{\mu} h)^2 
- \frac{Z_\gth X_0 M_0^{\prime}}{2}\, \lh \der_{\mu} \gth 
 + g \sqrt{\frac{2}{Z_{\gth}}}\, A_{\mu} \rh^2  }\\
 & & \\ 
 & & \dsp{ -\frac{g^2}{2}\, M_0^2 - \mu^2 X_0 - 
 \frac{g^2 X_0}{M_0^{\prime}}\, \lh M_0 M_0^{\prime\prime} 
 + M_0^{\prime\, 2} \rh h^2 + ... }\\
 & & \\
 & = & \dsp{ - m_W^2 W^+ \cdot W^- 
 - \frac{m_A^2}{2}\, \tilde{A}_{\mu}^2  - V_0 
 - \frac{1}{2}\, (\der_{\mu} h)^2  - \frac{m_h^2}{2}\, h^2 + ..., } 
\earr
\label{5.5}
\feq 
where we have only written out the quadratic terms which determine the linearized
field equations for the fluctuating part of the fields. To get the final result, 
we have taken 
\beq 
Z_h = \frac{1}{M_0^{\prime}}, \hs{2} Z_{\gth} = \frac{1}{X_0 M_0^{\prime}}, 
\label{5.6}
\feq
and 
\beq
m_W^2  = g^2 M_0, \hs{2} m_A^2 = 2 g^2 X_0 M_0^{\prime}, 
 \hs{2} m_h^2 = \frac{2g^2 X_0}{M_0^{\prime}}\, 
 \lh M_0 M_0^{\prime\prime} + M_0^{\prime\, 2} \rh, 
\label{5.7}
\feq
Also we have redefined the abelian vector field to absorb the Goldstone mode
in the usual way:
\beq
A_{\mu} \rightarrow \tilde{A}_{\mu} = A_{\mu} + \frac{1}{g}\, 
 \sqrt{\frac{Z_{\gth}}{2}}\, \der_{\mu} \gth.
\label{5.8}
\feq
This is equivalent to the choice of unitary gauge for the broken $U(1)$ symmetry. 
Finally, as $M_0$ and $M_0^{\prime}$ are related by (\ref{5.3}), we can 
eliminate $M_0$, $M_0^{\prime}$ and $M_0^{\prime\prime}$ and the 
$U(1)$ breaking parameter $X_0$ in favor of the physical parameters $m_W^2$, 
$m_A^2$, $m_h^2$ and the soft supersymmetry breaking parameter $\mu^2$: 
\beq 
g^2 M_0 = m_W^2, \hs{2} M_0^{\prime} = \frac{\mu^2}{m_W^2}, 
\hs{2} X_0 = \frac{m_A^2 m_W^2}{2 g^2 \mu^2}, 
\hs{2} M_0^{\prime\prime} = \frac{g^2 \mu^4}{m_A^2 m_W^6}\, 
 \lh m_h^2 - m_A^2 \rh.
\label{5.9}
\feq 
For values $\mu^2 > 0$ the model obviously describes massive charged and
neutral vector bosons plus a massive real Higgs scalar $h$. 

In the limit $\mu^2 \rightarrow 0$ we can now distinguish various 
possible scenarios: \nl
a.\ If there is a number $n > 0$ such that for small $\mu$ values 
$X_0 \sim \mu^{2n}$, then the $U(1)$ symmetry is restored when the
regulator mass vanishes; it also follows, that $m_A^2 m_W^2 \sim 
\mu^{2(n+1)}$. As $X_0 \rightarrow 0$ implies that  $M_0 \rightarrow 1/2$,
we find in this limit that $m_W^2 = g^2/2$ is finite and non-zero. Therefore 
$M^{\prime}(0) \sim \mu^2$, and $m_A^2 \sim \mu^{2(n+1)}$, which vanishes 
in the limit $\mu^2 \rightarrow 0$ as expected when $U(1)$ is restored. The 
last relation (\ref{5.9}) finally implies, that 
\beq 
m_h^2 \sim \mu^{2(n-1)} M^{\prime\prime}(0). 
\label{5.10}
\feq
For finite $M^{\prime\prime}(0) < \infty$, the scalar mass then remains finite 
for $n = 1$, and vanishes for $n > 1$. For $n < 1$ the scalar mass diverges. If 
$M^{\prime\prime}(0)$ itself vanishes as $\mu^p$, $p > 0$, these constraints 
can be further relaxed. The upshot is, that if $X_0$ vanishes at least as $\mu^2$, 
then in the limit $\mu^2 \rightarrow 0$ we reobtain the results of section 4. \nl
b.\ If in the supersymmetric limit of vanishing $\mu^2$ the vacuum expectation 
value  $X_0 > 0$, then the $U(1)$ symmetry remains broken. However, in the 
standard scenario with $m^2_W > 0$  the $U(1)$ gauge boson is massless in 
the limit $\gm^2 \rightarrow 0$. This apparent contradiction is resolved by 
looking at the kinetic term for the Goldstone field: it turns out that in the 
limit its effective charge 
\[
g_{eff} = g \sqrt{\frac{2}{Z_{\gth}}} = m_A \rightarrow 0.
\]
Therefore in this case there is a decoupling: the $U(1)$ symmetry broken
by $X_0$ is a global one, while the $U(1)$ gauge symmetry remains unbroken. 
Finally the $h$-scalar mass becomes  
\beq
m_h^2 \sim \frac{M_0^{\prime\prime}}{\mu^2}. 
\label{5.11}
\feq
Thus the scalar mass diverges, unless $M_0^{\prime\prime} \sim 
\mu^p$ with $p \geq 2$. 

In contrast, if there is a number $k > 0$ such that for small $\mu$ 
values $M_0 \sim \mu^{2k}$, then $m_W^2 \rightarrow 0$, and $m_A^2 
\sim \mu^{2(1 - k)}$. In this case one can not trust the limit $\mu^2 
\rightarrow 0$ to describe the critical $CP^1$-model, as the restauration 
of the $SU(2)$ symmetry is expected to be accompanied by the reappearance 
of light bosons $(z,\bz)$, and it is no longer allowed to use the unitary gauge. 

Nevertheless for finite $\mu^2$ the regularized model is well-defined 
and the mass spectrum can be computed. First of all, $m_A^2$ becomes 
large for small $\mu$ when $k > 1$; it is finite and $\mu$-independent for 
$k = 1$, and it vanishes for $k < 1$. Therefore the limit is well-behaved if 
$k \leq 1$. In that case the Higgs mass behaves as 
\beq
m_h^2 - m_A^2 \sim \mu^{2 (2k - 1)} M_0^{\prime\prime}. 
\label{5.12}
\feq
For finite $M_0^{\prime\prime}$ this implies that the masses remain finite
if $1/2 \leq k \leq 1$. In particular, for $k = 1$ we have $m_h^2  = m_A^2$,
both non-zero and finite. For $k = 1/2$ $m_h^2$ can be finite non-zero whilst
$m_A^2 = 0$. 

Turning to the fermion sector, the quadratic part of the lagrangean is
\beq 
\barr{lll}
\cL_{ferm} & = & \dsp{ - 2 M_0 \bgps_L \stackrel{\leftrightarrow}\sder \gps_L 
 - M_0^{\prime} \bgvf_L \stackrel{\leftrightarrow}{\sder} \gvf_L 
 - \bgl_L  \stackrel{\leftrightarrow}{\sder} \gl_L 
 - \bgl^+_L  \stackrel{\leftrightarrow}{\sder} \gl_L^- 
 - \bgl^-_L  \stackrel{\leftrightarrow}{\sder} \gl_L^+ }\\
 & & \\
 & & \dsp{ + 4g M_0 \lh \bgl_R^- \gps_L + \bgps_L \gl^+_R \rh 
 + 2 i g \sqrt{2 X_0}\, M^{\prime}_0 \lh \bgl_R \gvf_L - \bgvf_L \gl_R \rh. }
\earr
\label{5.13}
\feq
This is diagonalized by defining the Dirac spinors
\beq
\Psi = \sqrt{2M_0}\, \gps_L + \gl^+_R, \hs{2} 
\gF = \sqrt{M_0^{\prime}}\, \gvf_L - i \gl_R.
\label{5.14}
\feq
In terms of these fields, the expression (\ref{5.13}) becomes
\beq
\cL_{ferm} = - \bgPs  \stackrel{\leftrightarrow}{\sder} \gPs 
 - \bgF  \stackrel{\leftrightarrow}{\sder} \gF 
 - \bgl^+_L  \stackrel{\leftrightarrow}{\sder} \gl^-_L
 + 2g \sqrt{2M_0}\, \bgPs \gPs + 2g\sqrt{2X_0 M^{\prime}_0}\, \bgF \gF.
\label{5.15}
\feq 
It follows, that we have the fermion mass spectrum as given in table 2: 

\begin{center}
~~~~~~~~~~~~~~~~\begin{tabular}{|l|c|c|c|} \hline 
mass &  $m_{\gPs}^2$ & $m_{\gF}^2$ & $m_{\gl^-}^2$ \\ \hline
value & $2 g^2 M_0$ & $2 g^2 X_0 M_0^{\prime}$ & 0  \\ \hline
\end{tabular} \nl

{\footnotesize{Table 2: Fermion mass spectrum in the presence of
 soft supersymmetry breaking.}} 
\end{center} 

\nit
Combining the boson and fermion mass spectra we obtain
a supertrace formula including soft supersymmetry breaking:
\beq
\mbox{Str}\, m^2 = - \frac{2g^2 M_0}{M_0^{\prime}}\, \lh M^{\prime}_0
 - X_0 M^{\prime\prime}_0 \rh = -2 m_W^2 - m_A^2 + m_h^2. 
\label{5.16}
\feq
In particular, if in the limit $\mu^2 \rightarrow 0$ the $U(1)$ symmetry 
remains broken: $X_0 > 0$, and if in this limit the model is well-behaved, 
there are numbers $\omega^2 > 0$ and $1/2 \leq k \leq 1$ such that  for
small $\mu^2$ to first approximation
\beq
\barr{l} 
\dsp{ m_W^2=  \omega^2 \mu^{2k}, 
 \hs{2} m_A^2 = \frac{2g^2 X_0}{\omega^2}\,\mu^{2(1-k)}, }\\
 \\
\dsp{ m_h^2 = \frac{2g^2 X_0}{\omega^2}\, \mu^{2(1-k)} \lh 1 + 
 \frac{\omega^6}{g^2} \mu^{6k - 4} M^{\prime\prime}_0 \rh. }
\earr
\label{5.17}
\feq
Then
\beq
\mbox{Str}\, m^2 = -2 \omega^2 \mu^{2k} + 2 \omega^4 X_0 M_0^{\prime\prime}\,
 \mu^{2(2k-1)}\, 
 \stackrel{\mu^2 \rightarrow 0}{\longrightarrow}\,
 \left\{ \barr{ll} 0, &\mbox{if }\; 1/2 < k \leq 1; \\
         2 \omega^4 X_0 M_0^{\prime\prime}, & \mbox{if }\; k = 1/2. \earr
 \right. 
\label{5.18}
\feq
In fact, the supertrace vanishes even for $k >1$, but that is because 
the difference $m_h^2 - m_A^2$ then vanishes, even though both 
masses diverge individually.  \nl

\nit
The results (\ref{5.16}) and (\ref{5.18}) can be compared with standard
results for the supertrace formula \ct{cfgp,grisetal,GGRS}. This 
provides an excellent check on our results, as in the general form the 
supertrace of the mass matrix is computed in a gauge-independent way. 
Specifically, from the general lagrangean (\ref{b.2}) and (\ref{b.1}) 
we obtain
\beq
\mbox{Str}\, m^2 = \mbox{tr} \lh m_0^2 - 2m_{1/2}^2 + 3 m_1^2 \rh
\labl{5.19}
\feq
with the traces of the mass matrices for the various spins given by 
\beq
\mbox{tr}\, m_1^2 = 2 g^2 \lh M_0 + X_0 M^{\prime}_0 \rh, 
\label{5.20}
\feq
for the vectorbosons; 
\beq
\mbox{tr}\, m_{1/2}^2 = 4 g^2 \lh M_0 + X_0 M^{\prime}_0 \rh,
\label{5.21}
\feq
for the fermions; and finally
\beq
\mbox{tr}\, m_0^2 = 2G^{\uI I} V_{,I\uI} 
 = \frac{2(M_0 + X_0 M_0^{\prime})}{M_0 M_0^{\prime}}\, V^{\prime}_0
 + \frac{2X_0 V^{\prime\prime}_0}{M^{\prime}_0}.
\label{5.22}
\feq
To obtain this result, we have used the general expression for  
$G^{\uI I}$ in eq.(\ref{a.0}) in appendix A. Now observing that 
by definition of the vacuum state $X_0 V^{\prime}_0 = 0$, and that
\beq
\frac{2X_0 V^{\prime\prime}_0}{M^{\prime}_0}\, = 
 \frac{2g^2 X_0}{M^{\prime}_0} \lh M_0 M^{\prime\prime}_0 + 
 M^{\prime\,2}_0 \rh,
\label{5.23}
\feq
the final expression for the supertrace of the mass matrix
takes the form (\ref{5.16}): 
\beq
\mbox{Str}\, m^2 = \frac{2g^2M_0}{M^{\prime}_0}
  \lh X_0 M^{\prime\prime}_0 - M^{\prime}_0 \rh.
\label{5.24}
\feq
The regulator mass $\mu^2$ does not appear explicitly in this 
expression, because it does not contribute to $V^{\prime\prime}$. 
Observe, that the result (\ref{5.24}) even holds for $X_0 = 0$, 
due to the equality $g^2 M_0 = \mu^2/M_0^{\prime}$. Finally
observe, that in this derivation we have not used the unitary 
gauge at all.
\nl

\nit
{\bf 6.\ Examples} 
\vs{1} 

\nit
In  this section we provide examples of models with the properties conjectured in
sections 4 and 5. \nl
1.\ Let
\beq
K_m(X) = \gk_1 X + \frac{\gk_2}{2}\,  X^2 
\label{6.1}
\feq
Then 
\beq
M(X) = \frac{1}{2}\, + \gk_1 X + \gk_2 X^2, \hs{1.5} 
M^{\prime}(X) = \gk_1 + 2 \gk_2 X, \hs{1.5} M^{\prime\prime}(X) = 2 \gk_2. 
\label{6.2}
\feq
1.a If $\gk_1 > 0 $ and $\gk_2 \geq 0$, then $M(X)$  and $M^{\prime}(X)$ have 
no zeros for $X \geq 0$, and the potential reaches its minimum for $X = 0$; 
it follows that the $U(1)$ symmetry is preserved, whilst supersymmetry is broken 
as described in section 4, eqs.\ (\ref{3.6}) and (\ref{3.7}). \nl
1.b If $\gk_1 > 0$ and $\gk_2 < 0$, then $M(X)$ possesses a zero for 
\beq
X_0 = \frac{1}{ \gk_1 \lh \sqrt{1 +\frac{2|\gk_2|}{\gk_1^2}} -1 \rh}. 
\label{6.3}
\feq
However, if we include  the soft breaking term (\ref{5.1}),  we find that at the 
minimum $M < 0$ and $M^{\prime} < 0$, and the latter condition remains
true in the limit $\mu^2 \rightarrow 0$; hence the model contains
tachyons. We do not consider this case further. \nl 
1.c If $\gk_1 < 0$ and $\gk_2 >0$ with $\gk_1^2 < 2\gk_2$, then $M(X)$ has no 
zeros; however, $M^{\prime}(X) = 0$ at 
\beq
X_0 = \frac{|\gk _1|}{2\gk_2} < 1 \hs{1} \Rightarrow \hs{1} 
 M_0 = \frac{1}{2} \lh 1 - \frac{\gk_1^2}{2\gk_2} \rh.
\label{6.7.2}
\feq
This is the absolute minimum of the $D$-term potential. If we now include 
the soft supersymmetry breaking term with small $\mu^2$, we have to first 
approximation 
\beq
X_1 = X_0 + \gD X, \hs{2} M_1 = M_0, \hs{2} 
M_1^{\prime} = 2 \gk_2 \gD X,
\label{6.7.3}
\feq
with
\beq
\gD X = \frac{\mu^2}{2\gk_2 g^2 M_0}\, 
 = \frac{2\mu^2}{g^2 \lh 2\gk_2 - \gk_1^2 \rh}. 
\label{6.7.4}
\feq
In this case the mass spectrum of bosons reads to first approximation 
\beq
m_W^2 = \frac{g^2}{2} \lh 1  - \frac{\gk_1^2}{2\gk_2} \rh, \hs{1}
m_A^2 = \frac{4\mu^2 |\gk_1|}{2\gk_2 - \gk_1^2}, \hs{2}
m_h^2 = \frac{g^4 |\gk_1|}{2\mu^2} \lh 1 - \frac{\gk_1^2}{2\gk_2} \rh^2.
\label{6.7.5}
\feq
For the fermions we obtain the masses
\beq
m_{\gPs}^2 =2 m_W^2 = g^2 \lh 1  - \frac{\gk_1^2}{2\gk_2} \rh, \hs{2}
m_{\gF}^2 = m_A^2 = \frac{2\mu^2}{2\gk_2 - \gk_1^2}, \hs{2}
m_{\gl^-} = 0. 
\label{6.7.6}
\feq
1.d If $\gk_1 < 0$ and $\gk_2 > 0$ with $\gk_1^2 > 2\gk_2$, then 
$M(X)$ has two zeros, at 
\beq
X_{\pm} = \frac{1}{|\gk_1| \lh 1 \pm \sqrt{ 1 - \frac{2\gk_2}{\gk_1^2}} \rh}.
\label{6.7.7}
\feq
Again, including the soft breaking term the model contains tachyons at $X_+$. 
At $X_-$ it is well-behaved, with 
\beq
M_- = 0, \hs{2}M^{\prime}_- = |\gk_1| \sqrt{ 1 - \frac{2\gk_2}{\gk_1^2}}.
\label{6.7.8}
\feq
The physical minimum of the potential now occurs at 
\beq
X_1 = X_- + \gD X, 
\label{6.7.9}
\feq
where to first approximation in $\mu^2$ 
\beq
\gD X = \frac{\mu^2}{g^2 M_-^{\prime\, 2}} , \hs{2} 
M_1 = M^{\prime}_- \gD X, \hs{2} M^{\prime}_1 = M^{\prime}_-.
\label{6.1.1}
\feq
It follows, that the bosonic mass spectrum in this approximation reads
\beq
m_W^2 = \frac{\mu^2}{M_-^{\prime}}, \hs{2}
m_h^2 \approx m_A^2 = 2g^2 X_- M_-^{\prime} =  
 2g^2 \frac{\sqrt{1 - \frac{2\gk_2}{\gk_1^2}}}{1 - 
 \sqrt{1 - \frac{2\gk_2}{\gk_1^2}}}.
\label{6.1.2}
\feq
This corresponds to the results (\ref{5.17}) with $k =1$ and $\omega^2 = 
1/M_-^{\prime}$. The fermionic mass spectrum becomes 
\beq
m_{\gPs}^2 =  2 m_W^2 = \frac{2\mu^2}{M_-^{\prime}}, \hs{2} 
 m_{\gF}^2 = m_A^2 = 2 g^2 X_- M_-^{\prime},  \hs{2} m_{\gl^-}^2 = 0. 
\label{6.1.3}
\feq 
2.\ The above analysis is typical for models which do not have 
simultaneous zeros of $M(X)$ and its derivative $M^{\prime}(X)$.
If such simultanous zeros exist, as in the above model with $2\gk_2 = 
\gk_1^2$, the analysis is changed. As a generic example, consider the model
\beq
M(X) = \frac{1}{2}\, \lh \gk X - 1\rh^n, \hs{2} 
M^{\prime}(X) = \frac{n \gk}{2} \lh \gk X - 1\rh^{n-1}.
\label{6.2.1}
\feq 
When $n$ is a positive integer, an $n$-fold zero of $M(X)$ occurs at 
$X = 1/\gk$; it is also an $(n - 1)$-fold zero of $M^{\prime}(X)$. 
The minimum of the potential, including the soft breaking term, is at 
\beq
\frac{n \gk g^2}{4}\, \lh \gk X - 1 \rh^{2n-1} = \mu^2 \hs{1} 
 \Rightarrow \hs{1} \lh \gk X - 1 \rh = 
 \lh \frac{4\mu^2}{n \gk g^2} \rh^{\frac{1}{2n-1}}. 
\label{6.2.2}
\feq
It follows that at the minimum to lowest order in $\mu^2$:
\beq
\gk X_1 = 1 + \lh \frac{4\mu^2}{n\gk g^2} \rh^{\frac{1}{2n-1}}, \hs{2} 
M_1 = \frac{1}{2}\, \lh \frac{4\mu^2}{n\gk g^2} \rh^{\frac{n}{2n-1}}, 
 \hs{2} M_1^{\prime} = 
 \frac{n\gk}{2}\, \lh \frac{4\mu^2}{n\gk g^2} \rh^{\frac{n-1}{2n-1}}.
\label{6.2.3}
\feq    
Then the spectrum of boson masses becomes
\beq
m_W^2 = \frac{g^2}{2}\, 
\lh  \frac{4\mu^2}{n\gk g^2} \rh^{\frac{n}{2n-1}},  \hs{2} 
m_A^2 = n g^2 \lh \frac{4\mu^2}{n\gk g^2} \rh^{\frac{n-1}{2n-1}},
\label{6.2.4}
\feq
with the Higgs mass to lowest order in $\mu^2$: 
\beq 
\barr{ll}
\dsp{ m_h^2 = \frac{3}{2}\, m_A^2, }  & n = 2, \\
 & \\
m_h^2 = m_A^2, & n > 2. 
\earr
\label{6.2.5}
\feq
In all expressions we have kept only the terms of leading order in $\mu^2$
for small $\mu^2$. The fermion mass spectrum for these models reads
\beq 
m_{\gPs}^2 = g^2 \lh \frac{4\mu^2}{n\gk g^2} \rh^{\frac{n}{2n-1}}, \hs{2}
m_{\gF}^2 = ng^2 \lh \frac{4\mu^2}{n\gk g^2} \rh^{\frac{n-1}{2n-1}}, \hs{2}
m^2_{\gl^-} = 0. 
\label{6.2.6}
\feq
We observe, that in the limit $\mu^2 \rightarrow 0$ and for $n$ a positive
integer, all masses vanish, even though $X_0 = 1/\gk$ remains finite and 
non-zero. For $n = 1$ the masses $m_A^2 = m_{\gF}^2$ are finite non-zero, 
whilst for $1/2 < n < 1$ they diverge. 
\nl

\nit
{\bf 7.\ Discussion} 
\vs{1} 

\nit
In this paper we have investigated non-linear $\gs$-models with singular metrics, 
such that the kinetic terms of some fields vanish. Such models are for example 
supplied by gauged supersymmetric extensions of well-known coset-models. 

We have shown by general arguments and by regularization based upon 
the addition of soft supersymmetry breaking terms to the potential, that 
different types of behavior are possible. For example, the subset of linear
gauge symmetries can be realized manifestly, or in a spontaneously broken
mode; this is reflected in the mass of the corresponding vector bosons.  

In some cases the singularities imply the vanishing of all vector boson 
masses, which one expects to be accompanied by the reappearance of 
light bosons in the physical spectrum. However, this is difficult to show 
while staying in the original framework, as in particular it invalidates the
use of the unitary gauge. Of course, one could abandon the present 
approach and return to ordinary Yang-Mills theories with matter in linear 
representations; however, a more sophisticated formulation of the present 
models is possible and may shed light on this issue \ct{sgn-fr}. 

Our treatment of the $\gs$-model is based firmly on the classical action, 
although some features of our model were motivated by quantum aspects 
like the absence of holonomy and gauge anomalies. Higher order quantum 
corrections \ct{brig} may change the behavior of the models by renormalization 
of the K\"{a}hler potential. 

Another extension of interest would be to study what happens if one only 
gauges the linear stability group, i.e.\  $U(1)$ in the $CP^1$-model. This 
changes the $D$-terms and allows for the introduction of a Fayet-Iliopoulos 
term. Therefore it is likely that such models offer less problems to obtain physically 
reasonable spectra of masses. In the present paper we have not analyzed this 
modification. 

Finally, we have not studied in detail the issue of the appearance of 
tachyons in certain parameter ranges. The standard lore is that 
in this range the model is inconsistent. 
\vs{2} 

\nit
{\bf Acknowledgements} \nl
For three of us (T.S.N., F.R.\ and J.W.v.H) this work is part of the
research programme Theoretical Subatomic Physics (FP52) of the 
Foundation for Fundamental Research of Matter (FOM). S.G.N.\
acknowledges the support of CITA and NSERC.

\np

\nit
{\bf Appendix A}\nl

\nit
In this appendix we collect some results for the K\"{a}hler metric and its 
derivatives. The metric is given in (\ref{9}):  
\[
\barr{lll} 
G_{I\uI} & = & \dsp{  \frac{\der^2 K}{\der Z^I \der\bZ^{\uI}} 
 = \lh \barr{cc} G_{z\bz} & G_{z\ba} \\ 
                 G_{a\bz} & G_{a\ba} \earr \rh }\\ 
 \\ 
 & = & \dsp{ \lh \barr{cc} 
 \frac{1 + 2 (1 + 2 \bz z) X K_m^{\prime} 
 + 4 \bz z X^2 K_m^{\prime\prime} }{(1 + \bz z)^2}\, 
  &  2 \bz a (1 + \bz z)\, \lh 
 K_m^{\prime} + X K_m^{\prime\prime} \rh  \\
 & \\
 2 \ba z (1 + \bz z)\, \lh K_m^{\prime} + X K_m^{\prime\prime} \rh & 
 (1 + \bz z)^2 \lh K_m^{\prime} + X K_m^{\prime\prime} \rh 
       \earr \rh, }
\earr 
\]
with the inverse 
\beq
\barr{lll}
G^{\uI I} & = & \dsp{ \lh \barr{cc} 
  \frac{(1 + \bz z)^2}{1 + 2XK_m^{\prime}} & 
  - \frac{2 \bz a (1 + \bz z)}{1 + 2X K_m^{\prime}} \\
  & \\ 
  - \frac{2 \ba z (1 + \bz z)}{1 + 2X K_m^{\prime}} & 
  \frac{1}{(1 + \bz z)^2}\, \lh \frac{1}{K_m^{\prime} + X K_m^{\prime\prime}}
  + \frac{4\bz z X}{1 + 2X K_m^{\prime}} \rh \earr \rh. }
\earr 
\label{a.0}
\feq
We define the differential operators 
\beq
\der = dZ^I \dd{}{Z^I} = dz \dd{}{z} + da \dd{}{a}, \hs{2} 
\bar{\der} = \der \bZ^I \dd{}{\bZ^I} = d\bz \dd{}{\bz} + d\ba \dd{}{\ba}, 
\label{a.1}
\feq 
such that $d  = \der + \bar{\der}$. Their action on the variable $X$ is
given by 
\beq 
\der X = \frac{2 \bz dz}{1 + \bz z}\, X + \ba da (1 + \bz z)^2, \hs{2} 
\bar{\der} X = \frac{2 z d\bz}{1 + \bz z}\, X + a d\ba (1 + \bz z)^2.
\label{a.2} 
\feq
Applying these differential operators to the metric (\ref{9}) we get 
\beq 
\barr{lll}
\der G_{z\bz} & = & \dsp{ \frac{-2\bz dz}{(1 + \bz z)^3}\, \left[ 1 - 
 2 (1 + \bz z) X K_m^{\prime} - (4 + 10 \bz z) X^2 K_m^{\prime\prime} 
 - 4 \bz z X^3 K_m^{\prime\prime\prime} \right] }\\
 & & \\
 & & \dsp{ + 2 \ba da \left[ (1 + 2 \bz z) K_m^{\prime} + (1 + 6 \bz z) X 
 K_m^{\prime\prime} + 2 \bz z X^2 K_m^{\prime\prime\prime} \right]. }\\ 
 & & \\
\der G_{z\ba} & = & \dsp{ 2 a \bz^2 dz \left[ K_m^{\prime} + 
 5 X K_m^{\prime\prime} + 2 X^2 K_m^{\prime\prime\prime} \right] }\\
 & & \\
 & & \dsp{ + 2 \bz da (1 + \bz z) \left[ K_m^{\prime} + 3 X 
 K_m^{\prime\prime} + X^2 K_m^{\prime\prime\prime} \right]. }\\
 & & \\
\der G_{a\bz} & = & \dsp{ 2 \ba dz \left[ (1 + 2 \bz z) K_m^{\prime} 
 + (1 + 6 \bz z) X K_m^{\prime\prime} + 2 \bz z X^2 K_m^{\prime\prime\prime}
 \right] }\\  
 & & \\
 & & \dsp{ + 2 z \ba^2 da (1 + \bz z)^3 \left[ 2 K_m^{\prime\prime}
 + X K_m^{\prime\prime\prime} \right] }\\
 & & \\
\der G_{a\ba} & = & \dsp{ 2 \bz dz (1 + \bz z) \left[ K_m^{\prime} + 
 3 X K_m^{\prime\prime} + X^2 K_m^{\prime\prime\prime} \right] }\\
 & & \\ 
 & & \dsp{ + \ba da (1 + \bz z)^4 \left[ 2 K_m^{\prime\prime} + 
 X K_m^{\prime\prime\prime} \right], }
\earr
\label{a.3}
\feq 
and their complex conjugates. 
Next we compute the mixed second derivative of the metric components: 
\beq 
\bar{\der} \der G_{I\uI} = d\bz dz\, G_{I\uI,z\bz} + 
 d\ba dz\, G_{I\uI,z\ba} + d\bz da\, G_{I\uI,a\bz} + d\ba da\, G_{I\uI,a\ba}.
\label{a.4}
\feq 
This gives the following results: 
\beq
\barr{lll} 
G_{z\bz,z\bz} & = & \hs{-.5} \dsp{ \frac{1}{(1 + \bz z)^{4}}  \left[ -2 + 4 \bz z 
 + 4 (1 + \bz z)^2 X K_m^{\prime} + ( 8 + 64 \bz z + 68 (\bz z)^2 ) X^2 
 K_m^{\prime\prime} \rd }\\
 & & \\
 & & \hs{-.5} \dsp{ \ld +\, 16 \bz z (2 + 5 \bz z) X^3 K_m^{\prime\prime\prime} + 
 16 (\bz z)^2 X^4 K_m^{\prime\prime\prime\prime} \right] }\\
 & & \\
G_{z\bz,a\bz} & = & \hs{-.5} \dsp{ G_{a\bz,z\bz}\, =\, [ G_{z\bz,z\ba} ]^*\, =\, 
 [ G_{z\ba,z\bz} ]^* }\\
 & & \\
 & = & \hs{-.5} \dsp{ \frac{4\ba z}{(1+\bz z)} \left[ 
 (1 + \bz z) K_m^{\prime} + (5 + 11 \bz z) X K_m^{\prime\prime}  
 + (2 + 11 \bz z) X^2 K_m^{\prime\prime\prime} \rd }\\
 & & \\
 & & \hs {-.5} \dsp{ \ld +\, 2 \bz z X^3 K_m^{\prime\prime\prime\prime} \right] }\\
 & & \\
G_{z\bz, a\ba} & = & \hs{-.5} \dsp{ G_{z\ba,a\bz}\, =\, G_{a\bz,z\ba}\, =\, 
 G_{a\ba,z\bz} }\\
 & & \\
 & = & \hs{-.5} \dsp{ 2(1 + 2 \bz z) K_m^{\prime} + 2(3 + 14 \bz z) X 
 K_m^{\prime\prime} 
 + 2 (1 + 12 \bz z) X^2 K_m^{\prime\prime\prime} 
 + 4 \bz z X^3 K_m^{\prime\prime\prime\prime} }\\
 & & \\ 
G_{z\ba,z\ba} & = & \dsp{ 2(a\bz)^2 (1 + \bz z)^2 \left[ 6 K_m^{\prime\prime} + 
 9 X K_m^{\prime\prime\prime} + 2 X^2 K_m^{\prime\prime\prime\prime} \right] }\\
 & & \\
G_{z\ba,a\ba} & = & \hs{-.5} \dsp{ G_{a\ba,z\ba}\, =\, [G_{a\bz,a\ba} ]^*\, =\, 
 [G_{a\ba,z\ba}]^* }\\
 & & \\
 & = & \hs{-.5} \dsp{ 2 \bz a (1 + \bz z)^3
 \left[ 4 K_m^{\prime\prime} + 5 X K_m^{\prime\prime\prime} + 
 X^2 K_m^{\prime\prime\prime} \right] }\\
 & & \\ 
G_{a\ba,a\ba} & = & \hs{-.5}  \dsp{ (1 + \bz z)^4 \left[ 2 K_m^{\prime\prime}
 + 4 X K_m^{\prime\prime\prime} + X^2 K_m^{\prime\prime\prime\prime}
 \right] }
\earr 
\label{a.5} 
\feq 
We can now compute the components of the K\"{a}hler connection and 
curvature; in the unitary gauge $z = \bz = 0$ the non-trivial ones 
read explicitly 
\beq 
\barr{l}
\gG_{IJ}^{\;\;\;K} = G^{\uK K} G_{I\uK,J} \hs{1} \rightarrow \hs{1} \\
 \\ 
\dsp{ \gG_{az}^{\;\;\;z} = 2\ba\, \frac{K_m^{\prime} + X K_m^{\prime\prime}}{
 1 + 2 X K_m^{\prime}}, \hs{1} 
\gG_{aa}^{\;\;\;a} = \ba\, \frac{2 K_m^{\prime\prime} 
 + X K_m^{\prime\prime\prime}}{K_m^{\prime} + X K_m^{\prime\prime}}, }
\earr 
\label{a.6}
\feq 
and their complex conjugates, for the connection; and for the curvature: 
\beq
\barr{l}
K_{I\uI J\uJ} = G_{I\uI,J\uJ} - G^{\uK K} G_{I\uK,J} G_{K\uI,\uJ} \hs{1}
 \rightarrow \\
 \\
K_{z\bz z\bz} \hs{.2} = -2 (1 - 2 X K_m^{\prime} - 4 X^2 K_m^{\prime\prime}), \\
 \\
K_{z\bz a\ba}\, = K_{a\ba z\bz} = K_{z\ba a\bz} = K_{a\bz z \ba} \\
 \\
 \hs{1.5} = \frac{2}{1 + 2 X K_m^{\prime}} \lh K_m^{\prime} + 3 X 
 K_m^{\prime\prime} + X^2 K_m^{\prime\prime\prime} + 2 X^2 K_m^{\prime}
 K_m^{\prime\prime} + 2 X^3 (K_m^{\prime} K_m^{\prime\prime\prime} 
 - K_m^{\prime\prime} K_m^{\prime\prime}) \rh, \\
 \\
K_{a\ba a\ba} \hs{.1} = \frac{1}{K_m^{\prime} + X K_m^{\prime\prime}}\, 
 \lh 2 K_m^{\prime} K_m^{\prime\prime} + 2 X (K_m^{\prime} 
 K_m^{\prime\prime\prime} - K_m^{\prime\prime} K_m^{\prime\prime}) 
 + X^2 K_m^{\prime} K_m^{\prime\prime\prime} \rd \\
 \\
\ld \hs{3.7} +\, X^3 ( K_m^{\prime\prime} K_m^{\prime\prime\prime\prime} - 
 K_m^{\prime\prime\prime\, 2}) \rh. 
\earr
\label{a.7}
\feq 

\nit
{\bf Appendix B}\nl

\nit
In the singular limit of supersymmetric $\gs$-models, the elimination of 
auxiliary fields is a delicate procedure. Therefore in this appendix we 
present the off-shell lagrangean for our models (\ref{6}). The Lagrange 
density for the gauged supersymmetric CP$^1$ model plus the chiral 
matter multiplet is given on the next page. The Lagrange density of the 
gauge multiplet is: 
\beq
\barr{lll}
\cL_{gauge} & \hs{-.5} = \hs{-.5} & \dsp{
 - \frac{1}{2}\, F^+(W) \cdot F^-(W) - \frac{1}{4}\, F^2(A) }\\ 
 & & \\
 & & \dsp{ - \bgl_L^- \stackrel{\leftrightarrow}{\sDer} \gl_L^+ 
 - \bgl_L^+ \stackrel{\leftrightarrow}{\sDer} \gl_L^-  
 - \bgl_L \stackrel{\leftrightarrow}{\sDer} \gl_L + D^+ D^- + \frac{1}{2}\, D^2
 + \xi D. }
\earr 
\label{b.2}
\feq 
Here $\xi$ is the Fayet-Iliopoulos parameter, which can only be included in 
the model with gauged linear $U(1)$ (and hence $W^{\pm} = 0$); furthermore 
the decomposition of the vector multiplet $V = (W_{\mu}, \gl, D)$ is defined by 
\beq 
V = V^i \tau_i, \hs{2} V^{\pm} = \frac{1}{\sqrt{2}}\, \lh V^1 \pm i V^2 \rh, \hs{2} 
 V = V^3. 
\label{b.3}
\feq 
In particular: $W^{\pm} = (W^1 \pm i W^2)/\sqrt{2}$, $A = W^3$, and
\beq 
\barr{lll}
F^{\pm}_{\mu\nu}(W) & = & \dsp{ \lh \der_{\mu} \mp ig A_{\mu} \rh W^{\pm}_{\nu} - 
 \lh \der_{\nu} \mp ig A_{\nu} \rh W^{\pm}_{\mu}, }\\
 & & \\ 
F_{\mu\nu}(A) & = & \dsp{ \der_{\mu} A_{\nu} - \der_{\nu} A_{\mu} - 
 ig W^+_{\mu} W^-_{\nu} + ig W^+_{\nu} W^-_{\mu}. }
\earr 
\label{b.4}
\feq 
Elimination of the auxiliary $D$-fields from (\ref{b.2}) with $\xi = 0$, and using 
the Lagrange density below, then leads to the potential $V_D$, eq.\ (\ref{13}). 

\beq
\barr{lll} 
\cL_{\gs + m} \hs{-.5} & = \hs{-.5}&  
 \dsp{ - G_{z\bz} \lh D\bz \cdot Dz + \bar{\psi}_L\stackrel{\leftrightarrow}{\sDer} 
 \psi_L  - \bH H \rh - G_{a\ba} \lh D\ba \cdot Da + \bar{\varphi}_L 
 \stackrel{\leftrightarrow}{\sDer} \varphi_L  - \bB B \rh }\\
 & & \\
 & & \dsp{ - G_{z\ba} \lh D\ba \cdot Dz + \bar{\varphi}_L 
 \stackrel{\leftrightarrow}{\sDer} \psi_L  - \bB H \rh 
 - G_{a\bz} \lh D\bz \cdot Da + \bar{\psi}_L 
 \stackrel{\leftrightarrow}{\sDer} \varphi_L  - \bH B \rh }\\
 & & \\
 & & \dsp{ - \lh G_{z\bz,z} D_{\mu}z - G_{z\bz, \bz} D_{\mu} \bz 
 + G_{z\bz,a} D_{\mu}a - G_{z\bz,\ba} D_{\mu} \ba \rh\, \bar{\psi}_L \gg^{\mu} 
 \psi_L }\\
 & & \\
 & & \dsp{ - \lh G_{a\ba,z} D_{\mu}z - G_{a\ba, \bz} D_{\mu} \bz 
 + G_{a\ba,a} D_{\mu}a - G_{a\ba,\ba} D_{\mu} \ba \rh\, \bar{\varphi}_L
  \gg^{\mu} \varphi_L }\\
 & & \\
 & & \dsp{ - \lh G_{z\ba,z} D_{\mu}z - G_{z\ba, \bz} D_{\mu} \bz 
 + G_{z\ba,a} D_{\mu}a - G_{z\ba,\ba} D_{\mu} \ba \rh\, \bar{\varphi}_L \gg^{\mu}
 \psi_L }\\
 & & \\
 & & \dsp{ - \lh G_{a\bz,z} D_{\mu}z - G_{a\bz, \bz} D_{\mu} \bz 
 + G_{a\bz,a} D_{\mu}a - G_{a\bz,\ba} D_{\mu} \ba \rh\, \bar{\psi}_L \gg^{\mu} 
 \varphi_L }\\
 & & \\
 & & \dsp{ - \lh G_{z\bz,z} \bH + G_{z\ba,z} \bB \rh\, \bar{\psi}_R \psi_L 
 - \lh G_{z\bz,\bz} H + G_{a\bz,\bz} B \rh\, \bar{\psi}_L \psi_R }\\ 
 & & \\
 & & \dsp{ - \lh G_{a\bz,a} \bH + G_{a\ba,a} \bB \rh\, \bar{\varphi}_R \varphi_L 
 - \lh G_{z\ba,\ba} H + G_{a\ba,\ba} B \rh\, \bar{\varphi}_L \varphi_R }\\ 
 & & \\ 
 & & \dsp{ -  2\lh G_{z\bz,a} \bH + G_{z\ba,a} \bB \rh\, \bar{\varphi}_R \psi_L 
 -  2\lh G_{z\bz,\ba} H + G_{a\bz,\ba} B \rh\, \bar{\varphi}_L \psi_R }\\ 
 & & \\
 & & \dsp{ +\,  G_{z\bz,z\bz} \bar{\psi}_R\psi_L \bar{\psi}_L \psi_R 
 +  2\, G_{z\bz,a\bz} \bar{\varphi}_R\psi_L \bar{\psi}_L \psi_R 
 +  2\, G_{z\bz,z\ba} \bar{\psi}_R\psi_L \bar{\psi}_L \varphi_R }\\
 & & \\
 & & \dsp{ +\, G_{z\ba,z\ba} \bar{\psi}_R\psi_L \bar{\varphi}_L \varphi_R 
 +  G_{a\bz,a\bz} \bar{\varphi}_R\varphi_L \bar{\psi}_L \psi_ R  
 + 4\,  G_{z\bz,a\ba} \bar{\varphi}_R\psi_L \bar{\psi}_L \varphi_ R  }\\ 
 & & \\
 & & \dsp{ +\,  G_{a\ba,a\ba} \bar{\varphi}_R\varphi_L \bar{\varphi}_L \varphi_R 
 +  2\, G_{a\ba,a\bz} \bar{\varphi}_R\varphi_L \bar{\varphi}_L \psi_R 
 +  2\, G_{a\ba,z\ba} \bar{\psi}_R\varphi_L \bar{\varphi}_L \varphi_R }\\
 & & \\
 & & \dsp{ +\, 2 g\, G_{z\bz} \left[ \lh \bgl_R^-  - i \sqrt{2}\, \bz \bgl_R
 + \bz^2 \bgl_R^+ \rh \psi_L  + \bar{\psi}_L \lh \gl_R^+ + i \sqrt{2}\, z \gl_R 
 + z^2 \gl_R^- \rh \right] }\\
 & & \\
 & & \dsp{ +\, 2 g\, G_{z\ba} \left[ \lh  i  \sqrt{2}\, \ba \bgl_R 
 - 2\bz \ba \bgl_R^+ \rh \psi_L + \bar{\varphi}_L \lh \gl_R^+ 
 + i \sqrt{2}\, z \gl_R + z^2 \gl_R^- \rh  \right] }\\ 
 & & \\
 & & \dsp{ +\, 2 g\, G_{a\bz} \left[ \lh \bgl_R^- - i \sqrt{2}\, \bz \bgl_R 
 + z^2 \bgl_R^+ \rh \varphi_L - \bar{\psi}_L \lh  i \sqrt{2}\, a \gl_R  
 + 2z a \gl_R^- \rh  \right] }\\ 
& & \\
 & & \dsp{ +\, 2 g\, G_{a\ba} \left[ \lh  i \sqrt{2}\, \ba \bgl_R 
 - 2\bz \ba \bgl_R^+ \rh \varphi_L  - \bar{\varphi}_L \lh  
 i \sqrt{2}\, a \gl_R  + 2z a \gl_R^- \rh  \right] }\\ 
 & & \\
 & & \dsp{ 
 -\, \frac{g}{2}\, \lh 1 + 2 X K^{\prime}_m(X) \rh\, \frac{ D (1 - \bz z) 
 + i \sqrt{2}\, (D^+ \bz - D^- z)}{1 + \bz z}. }
\earr 
\label{b.1} 
\feq 
In addition to the potential $V_D$ generated by the $D$-terms, we observe 
that the equations for the auxiliary fields $(H, B)$ and their complex conjugates 
become 
\beq 
\barr{l} 
G_{z\bz} H + G_{a\bz} B = G_{z\bz,z}\, \bar{\psi}_R \psi_L + G_{a\bz,a}\, 
 \bar{\varphi}_R \varphi_L + 2 G_{z\bz,a} \bar{\varphi}_R \psi_L, \\
 \\ 
G_{z\ba} H + G_{a\ba} B = G_{z\ba,z}\, \bar{\psi}_R \psi_L + G_{a\ba,a}\, 
 \bar{\varphi}_R \varphi_L  + 2 G_{z\ba,a} \bar{\varphi}_R \psi_L, 
\earr 
\label{b.5}
\feq 
and their conjugates. 

In the unitary gauge $z = \bz = 0$ this becomes 
\beq
\barr{lll} 
\cL_{\gs + m} \hs{-.5} & = \hs{-.5}&  \dsp{ - 2 M(X)  
 \lh \frac{g^2}{2}\, W^- \cdot W^+ + \bar{\psi}_L 
 \hs{-.2}\stackrel{\leftrightarrow}{\sDer} \hs{-.2} \psi_L  - \bH H \rh }\\
 & & \\
 & & \dsp{ - M^{\prime}(X) \lh D\ba \cdot Da + \bar{\varphi}_L 
 \hs{-.2} \stackrel{\leftrightarrow}{\sDer} \hs{-.2} \varphi_L  - \bB B \rh 
 -g M(X) D }\\
 & & \\
 & & \dsp{ -\, 2\, (\ba\hs{-.2}\stackrel{\leftrightarrow}{D}_{\mu}\hs{-.2}a)
 \left[ M^{\prime}(X)\, \bar{\psi}_L \gg^{\mu} \psi_L + 
 M^{\prime\prime}(X)\,  \bar{\varphi}_L \gg^{\mu} \varphi_L \right] }\\
 & & \\
 & & \dsp{ +\,  \sqrt{2}\,g M^{\prime}(X)\, \left[ 
 - a W^-_{\mu}\, \bar{\varphi}_L \gg^{\mu}  \psi_L  + \ba W^+_{\mu}\, 
 \bar{\psi}_L  \gg^{\mu} \varphi_L \right]   }\\ 
 & & \\ 
 & & \dsp{ -M^{\prime\prime}(X)\, 
 \lh \ba \bB\, \bgvf_R \gvf_L  + a B\, \bgvf_L \gvf_R \rh  }\\
 & & \\
 & & \dsp{ -\,  4 M^{\prime}(X)\, \lh \ba \bH\,  \bgvf_R \psi_L 
 + aH\, \bgvf_L \psi_R \rh }\\ 
 & & \\
 & & \dsp{ -\, 4  \lh M(X) - 2 X M^{\prime}(X) \rh
 \bar{\psi}_R \psi_L\, \bar{\psi}_L \psi_R }\\
 & & \\
 & & \dsp{ +\, \lh M^{\prime\prime}(X) + X M^{\prime\prime\prime}(X)  \rh  
 \bgvf_R \gvf_L\, \bgvf_L \gvf_R }\\
 & & \\
 & & \dsp{ +\, 8 \lh M^{\prime}(X) + X M^{\prime\prime}(X) \rh 
 \bgvf_R \psi_L\, \bar{\psi}_L \gvf_R }\\
 & & \\ 
 & & \dsp{ +\, 2 \sqrt{2}\, g\,  \left[ \sqrt{2}\, M(X) \lh \bgl_R^- \psi_L + 
 \bar{\psi}_L \gl_R^+ \rh \rd }\\
 & & \\
 & & \dsp{ \ld \hs{4.2} +\, i M^{\prime}(X)\,
 \lh \ba \bgl_R\, \gvf_L - a \bgvf_L \gl_R \rh \right] }
\earr
\label{b.6}
\feq 
In case the superpotential vanishes the dependence on the charged vector 
fields $W^{\pm}$ and their superpartners $\gl^{\pm}$ and auxiliary fields 
$D^{\pm}$ disappears, and the model is indistinguishable of that with gauged 
$U(1)$ only. 

In the unitary gauge the equations for the auxiliary fields become 
\beq 
\barr{l} 
M(X)\, H = 2 \ba M^{\prime}(X)\, \bgvf_R \psi_L, \\
 \\
M^{\prime}(X)\, B = \ba M^{\prime\prime}(X)\, \bgvf_R \gvf_L.
\earr
\label{b.7}
\feq

\end{document}